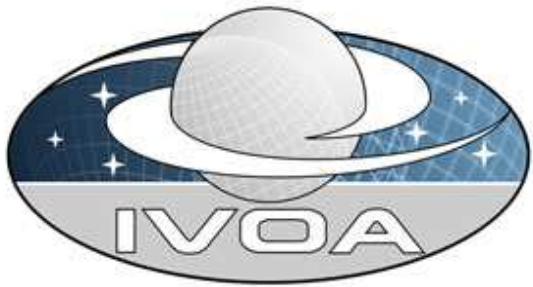

*International*

*Virtual*

*Observatory*

*Alliance*

# The UCD1+ controlled vocabulary Version 1.23

## *IVOA Recommendation 2007 April 2*

**This version:**
   http://www.ivoa.net/Documents/REC/UCD/UCDlist-20070402.html
**Latest version:**
   http://www.ivoa.net/Documents/latest/UCDlist.html
**Previous version(s):**
   http://www.ivoa.net/Documents/REC/UCD/UCDlist-20051231.html
**Editor(s):**
   A. Preite Martinez, S. Derriere
**Author(s):**

Andrea Preite Martinez (andrea.preitemartinez@iasf-roma.inaf.it),
Sebastien Derriere (derriere@astro.u-strasbg.fr),
Nausicaa Delmotte (ndelmot@eso.org),
Norman Gray (norman@astro.gla.ac.uk),
Robert Mann (rgm@roe.ac.uk),
Jonathan McDowell (jcm@cfa.harvard.edu),
Thomas Mc Glynn (Thomas.A.McGlynn@nasa.gov),
François Ochsenbein (francois@astro.u-strasbg.fr),
Pedro Osuna (Pedro.Osuna@esa.int),
Guy Rixon (gtr@ast.cam.ac.uk),
Roy Williams (roy@cacr.caltech.edu)

## Abstract

This document describes the list of controlled terms used to build the Unified Content Descriptors, Version 1+ (UCD1+). The document describing the UCD1+ can be found at the URL: http://www.ivoa.net/Documents/latest/UCD.html. This document reviews the structure of the UCD1+ and presents the current vocabulary.

## Status of This Document

*This is an IVOA Recommendation. This document has been produced by the IVOA Semantics Working Group. It has been reviewed by IVOA Members and other interested parties, and has been endorsed by the IVOA Executive Committee as an IVOA Recommendation. It is a stable document and may be used as reference material or cited as a normative reference from another document. IVOA's role in making the Recommendation is to draw attention to the specification and to promote its widespread deployment. This enhances the functionality and interoperability inside the Astronomical Community.*

*A list of current IVOA Recommendations and other technical documents can be found at http://www.ivoa.net/Documents/.*

## Acknowledgements

This document is based on the W3C documentation standards as adapted for the IVOA.

## Contents

# 1  Definition of atoms and words

A UCD is a string which contains textual tokens called 'words', separated by semicolons(;). A word is composed of 'atoms', separated by periods(.). The hierarchy is as follows:

                atoms --> words --> composed words

UCD1+ are either single words, or a composition of several words.

UCDs are "controlled" (through a process that is also indicated in the reference document above). Control is exercised at the level of words (UCD1+) and at the level of the vocabulary (atoms) used to form words. A consistent list of atoms will be mantained, making sure that the same atom always means the same thing, even if used in combination with different other atoms.

## 1.1  Definition of atoms

Atoms are defined following these guidelines:

1. Abbreviations are kept to a minimum, and only if the result is not ambiguous. (**ra**, **dec** are acceptable, but **t** is ambiguous: **time** and **temperature** are used instead.)

2. Atoms are not hyphenated. The separation is marked by a capital letter to help readability (position angle = **posAng**) unless the composed word has a well known acronym (signal to noise ratio = **snr**) or short form (standard deviation = **stdev**). There are only two exception to this rule: (i) the X-ray band (**em.X-ray**) and (ii) the frequency / wavelength intervals defining regions of the e.m. spectrum (e.g. **em.radio.3-6GHz**).

## 1.2  Definition of words

The list of UCD1+ words presented in this document was initially generated applying the rules and recommendations of PR-UCD-20040823 to catalogues/tables in VizieR. The original motivation was to transform old UCD1 into an improved version, trying to build a list of combinations of new words that could describe all the existing UCD1 terms.

The list of UCD1+ words is maintained by the UCD Scientific Board, following the procedure defined in the UCD Recommendation document (http://www.ivoa.net/Documents/latest/UCD.html), and described in detail in http://www.ivoa.net/Documents/latest/UCDlistMaintenance.html .

## 2  The structure of the UCD1+ tree

All existing UCD1+ words are grouped into 12 main categories. These categories are expressed by the first atom of the word, whose possible values are:

1. **arith** (arithmetics)

    This section includes concepts involving or indicating some mathematical operation performed on the primary 'concept' or just the presence of an arithmetic factor or operator.

2. **em** (electromagnetic spectrum)

    This section describes the electromagnetic spectrum, either in a monochromatic way or in predefined intervals. The complete list of proposed bands (in seven classical regions of the e.m. spectrum: radio, millimeter, infrared, optical, ultraviolet, x-ray and gamma-ray), can be found in the document [Note-EMSpectrum-20040520](Note-EMSpectrum-20040520)

3. **instr** (instrument)

    This section includes all quantities related to astronomical instrumentation, e.g. detectors (plates, CCDs, etc.), spectrographs, and telescopes (including observatories or missions), etc.

4. **meta** (metadata)

    This section includes all the information that is not coming directly from a measurement, and information that could not be included in other sections.

5. **obs** (observation)

    In principle under this section should go all words describing an observation (the name of the observer or PI, the observing conditions, the name of the field). In practice, the section is very 'thin' and could be deleted, if the sparse content could be housed elsewere.

6. **phot** (photometry)

    All the words describing photometric measures are included in this section. The definitions distinguish between a flux density (flux per unit frequency interval), a flux density integrated over a given e.m. interval (flux if expressed linearly, mag if expressed by a log), or a flux expressed in counts/s (if the setup of the detector is photon counting observing mode). 'Colors', which are differences of magnitudes (i.e. ratios of fluxes) measured in different bandpasses, are also included.

7. **phys** (physics)

    This section includes atomic and molecular data (mainly used for spectroscopy) and basic physical quantities (temperature, mass, gravity, luminosity, etc.)

8. **pos** (positional data)

    This section describes all quantities related to the position of an object on the sky:

    - Angular coordinates, and projections from spherical to rectangular systems.
    - Angular measurements in general (the angular size of an object is in this section, its linear size is in the **phys** section).
    - The WCS FITS keywords.

9. **spect** (spectral data)

    For historical reasons, photometric data taken in narrow spectral bands with instruments called spectrographs are classified as spectroscopic data. These definitions should not be confused with those in the **em** category. **em** represents the independent variable, or dispersion axis, and **phot** and **spect** describe the dependent variable, or flux axis.

10. **src** (source)

    This is a rather generic section, mainly devoted to source classifications. Variability, orbital, and velocity data are also included in this section.

11. **stat** (statistics)

    This section includes statistical information on measurements.

12. **time** (time)

    Quantities related to time (age, date, period, etc.) are described in this section.

## Appendix A: List of valid words

All words are preceded by a 'syntax' code that can help in the process of building composed UCD1+.

1. The code "P" means that the word can only be used as "primary" or first word;
2. "S" stands for only secondary: the word cannot be used as the first word to describe a single quantity;
3. "Q" means that the word can be used indifferently as first or secondary word;
4. "E" means a photometric quantity, and can be followed by a word describing a part of the electromagnetic spectrum
5. "C" is a colour index, and can be followed by two successive word describing a part of the electromagnetic spectrum;
6. "V" stands for vector. Such a word can be followed by another describing the axis or reference frame in which the measurement is done

| Code | UCD | Description |
|---|---|---|
| Q | arith | Arithmetic quantities |
| S | arith.diff | Difference between two quantities described by the same UCD |
| P | arith.factor | Numerical factor |
| P | arith.grad | Gradient |
| P | arith.rate | Rate (per time unit) |
| S | arith.ratio | Ratio between two quantities described by the same UCD |
| Q | arith.zp | Zero point |
| S | em | Electromagnetic spectrum |
| S | em.radio | Radio part of the spectrum |
| S | em.radio.20-100MHz | Radio between 20 and 100 MHz |
| S | em.radio.100-200MHz | Radio between 100 and 200 MHz |
| S | em.radio.200-400MHz | Radio between 200 and 400 MHz |
| S | em.radio.400-750MHz | Radio between 400 and 750 MHz |
| S | em.radio.750-1500MHz | Radio between 750 and 1500 MHz |
| S | em.radio.1500-3000MHz | Radio between 1500 and 3000 MHz |
| S | em.radio.3-6GHz | Radio between 3 and 6 GHz |
| S | em.radio.6-12GHz | Radio between 6 and 12 GHz |
| S | em.radio.12-30GHz | Radio between 12 and 30 GHz |
| S | em.mm | Millimetric part of the spectrum |
| S | em.mm.30-50GHz | Millimetric between 30 and 50 GHz |
| S | em.mm.50-100GHz | Millimetric between 50 and 100 GHz |
| S | em.mm.100-200GHz | Millimetric between 100 and 200 GHz |
| S | em.mm.200-400GHz | Millimetric between 200 and 400 GHz |
| S | em.mm.400-750GHz | Millimetric between 400 and 750 GHz |
| S | em.mm.750-1500GHz | Millimetric between 750 and 1500 GHz |
| S | em.mm.1500-3000GHz | Millimetric between 1500 and 3000 GHz |
| S | em.IR | Infrared part of the spectrum |
| S | em.IR.J | Infrared between 1.0 and 1.5 micron |
| S | em.IR.H | Infrared between 1.5 and 2 micron |
| S | em.IR.K | Infrared between 2 and 3 micron |
| S | em.IR.3-4um | Infrared between 3 and 4 micron |
| S | em.IR.4-8um | Infrared between 4 and 8 micron |
| S | em.IR.8-15um | Infrared between 8 and 15 micron |
| S | em.IR.15-30um | Infrared between 15 and 30 micron |
| S | em.IR.30-60um | Infrared between 30 and 60 micron |
| S | em.IR.60-100um | Infrared between 60 and 100 micron |
| S | em.IR.NIR | Near-Infrared, 1-5 microns |
| S | em.IR.MIR | Medium-Infrared, 5-30 microns |
| S | em.IR.FIR | Far-Infrared, 30-1000 microns |
| S | em.opt | Optical part of the spectrum |
| S | em.opt.U | Optical band between 300 and 400 nm |
| S | em.opt.B | Optical band between 400 and 500 nm |
| S | em.opt.V | Optical band between 500 and 600 nm |
| S | em.opt.R | Optical band between 600 and 750 nm |
| S | em.opt.I | Optical band between 750 and 1000 nm |
| S | em.UV | Ultraviolet part of the spectrum |
| S | em.UV.10-50nm | Ultraviolet between 10 and 50 nm |
| S | em.UV.50-100nm | Ultraviolet between 50 and 100 nm |
| S | em.UV.100-200nm | Ultraviolet between 100 and 200 nm |
| S | em.UV.200-300nm | Ultraviolet between 200 and 300 nm |
| S | em.UV.FUV | Far-Ultraviolet |
| S | em.X-ray | X-ray part of the spectrum |
| S | em.X-ray.soft | Soft X-ray (0.12 - 2 keV) |
| S | em.X-ray.medium | Medium X-ray (2 - 12 keV) |
| S | em.X-ray.hard | Hard X-ray (12 - 120 keV) |
| S | em.gamma | Gamma rays part of the spectrum |
| S | em.gamma.soft | Soft gamma ray (120 - 500 keV) |
| S | em.gamma.hard | Hard gamma ray (>500 keV) |
| S | em.line | Designation of major atomic lines |
| S | em.line.Brgamma | Bracket gamma line |
| S | em.line.HI | 21cm hydrogen line |
| S | em.line.Halpha | H-alpha line |
| S | em.line.Hbeta | H-beta line |
| S | em.line.Hgamma | H-gamma line |
| S | em.line.Hdelta | H-delta line |
| S | em.line.Lyalpha | Hydrogen Lyman alpha line |
| S | em.line.OIII | [OIII] line whose rest wl is 500.7 nm |
| S | em.line.CO | CO radio line, e.g. 12CO(1-0) at 115GHz |
| Q | em.bin | Channel / instrumental spectral bin coordinate (bin number) |
| Q | em.energy | Energy value in the em frame |
| Q | em.freq | Frequency value in the em frame |
| Q | em.wavenumber | Wavenumber value in the em frame |
| Q | em.wl | Wavelength value in the em frame |
| Q | em.wl.central | Central wavelength |
| Q | em.wl.effective | Effective wavelength |
| Q | instr | Instrument |
| E | instr.background | Instrumental background |
| Q | instr.bandpass | Bandpass (e.g.: band name) of instrument |
| Q | instr.bandwidth | Bandwidth of the instrument |
| Q | instr.baseline | Baseline for interferometry |
| S | instr.beam | Beam |
| Q | instr.calib | Calibration parameter |
| S | instr.det | Detector |
| Q | instr.det.noise | Instrument noise |
| Q | instr.det.psf | Point Spread Function |
| Q | instr.det.qe | Quantum efficiency |
| Q | instr.dispersion | Dispersion of a spectrograph |
| S | instr.filter | Filter |
| S | instr.fov | Field of view |
| S | instr.obsty | Observatory, satellite, mission |
| Q | instr.obsty.seeing | Seeing |
| Q | instr.offset | Offset angle respect to main direction of observation |
| Q | instr.order | Spectral order in a spectrograph |
| Q | instr.param | Various instrumental parameters |
| S | instr.pixel | Pixel (default size: angular) |
| S | instr.plate | Photographic plate |
| Q | instr.plate.emulsion | Plate emulsion |
| Q | instr.precision | Instrument precision |
| Q | instr.saturation | Instrument saturation threshold |
| Q | instr.scale | Instrument scale (for CCD, plate, image) |
| Q | instr.sensitivity | Instrument sensitivity, detection threshold |
| Q | instr.setup | Instrument configuration or setup |
| Q | instr.skyLevel | Sky level |
| Q | instr.skyTemp | Sky temperature |
| Q | instr.tel | Telescope |
| Q | instr.tel.focalLength | Telescope focal length |
| P | meta | Metadata |
| P | meta.abstract | Abstract (of paper, proposal,etc.) |
| P | meta.bib | Bibliographic reference |
| P | meta.bib.author | Author name |
| P | meta.bib.bibcode | Bibcode |
| P | meta.bib.fig | Figure in a paper |
| P | meta.bib.journal | Journal name |
| P | meta.bib.page | Page number |
| P | meta.bib.volume | Volume number |
| P | meta.code | Code or flag |
| P | meta.code.class | Classification code |
| P | meta.code.error | limit uncertainty error flag |
| P | meta.code.member | Membership code |
| P | meta.code.mime | MIME type |
| P | meta.code.multip | Multiplicity or binarity flag |
| P | meta.code.qual | Quality, precision, reliability flag or code |
| P | meta.code.status | Status code (e.g.: status of a proposal/observation) |
| P | meta.cryptic | Unknown or impossible to understand quantity |
| P | meta.curation | Identity of man/organization responsible for the data |
| Q | meta.dataset | Dataset, archive |
| Q | meta.email | Curation/contact e-mail |
| S | meta.file | File |
| S | meta.fits | FITS standard |
| P | meta.id | Identifier, name or designation |
| P | meta.id.assoc | Identifier of associated counterpart |
| P | meta.id.CoI | Name of Co-Investigator |
| P | meta.id.cross | Cross identification |
| P | meta.id.parent | Identification of parent source |
| P | meta.id.part | Part of identifier, suffix or sub-component |
| P | meta.id.PI | Name of Principal Investigator |
| S | meta.main | Main value of something |
| S | meta.modelled | Quantity was produced by a model |
| P | meta.note | Note or remark (longer than a code or flag) |
| P | meta.number | Number (of things; e.g. nb of object in an image) |
| P | meta.record | Record number |
| P | meta.ref | Reference, or origin |
| Q | meta.ref.ivorn | IVORN, Int. VO Resource Name (ivo://) |
| Q | meta.ref.uri | URI, universal resource identifier |
| P | meta.ref.url | URL, web address, service endpoint |
| S | meta.software | Software used in generating data |
| S | meta.table | Table or catalogue |
| P | meta.title | Title or explanation |
| Q | meta.ucd | UCD |
| P | meta.unit | Unit |
| P | meta.version | Version |
| S | obs | Observation |
| Q | obs.airMass | Airmass |
| S | obs.atmos | Atmosphere, atmospheric phenomena affecting an observation |
| Q | obs.atmos.extinction | Atmospheric extinction |
| Q | obs.atmos.refractAngle | Atmospheric refraction angle |
| S | obs.calib | Calibration observation |
| S | obs.calib.flat | Related to flat-field calibration observation (dome, sky, ..) |
| S | obs.exposure | Exposure |
| S | obs.field | Region covered by the observation |
| S | obs.image | Image |
| Q | obs.observer | Observer, discoverer |
| Q | obs.param | Various observation or reduction parameter |
| S | obs.proposal | Observation proposal |
| Q | obs.proposal.cycle | Proposal cycle |
| S | obs.sequence | Sequence of observations, exposures or events |
| E | phot | Photometry |
| E | phot.antennaTemp | Antenna temperature |
| Q | phot.calib | Photometric calibration |
| C | phot.color | Color index or magnitude difference |
| Q | phot.color.excess | Color excess |
| Q | phot.color.reddFree | Dereddened, reddening-free color |
| E | phot.count | Flux expressed in counts |
| E | phot.fluence | Fluence |
| E | phot.flux | Photon flux |
| Q | phot.flux.bol | Bolometric flux |
| E | phot.flux.density | Flux density (per wl/freq/energy interval) |
| E | phot.flux.density.sb | Flux density surface brightness |
| E | phot.flux.sb | Flux surface brightness |
| E | phot.limbDark | Limb-darkening coefficients |
| E | phot.mag | Photometric magnitude |
| Q | phot.mag.bc | Bolometric correction |
| Q | phot.mag.bol | Bolometric magnitude |
| Q | phot.mag.distMod | Distance modulus |
| E | phot.mag.reddFree | Dereddened magnitude |
| E | phot.mag.sb | Surface brightness in magnitude units |
| Q | phys | Physical quantities |
| Q | phys.SFR | Star formation rate |
| E | phys.absorption | Extinction or absorption along the line of sight |
| Q | phys.absorption.coeff | Absorption coefficient (e.g. in a spectral line) |
| Q | phys.absorption.gal | Galactic extinction |
| Q | phys.absorption.opticalDepth | Optical depth |
| Q | phys.abund | Abundance |
| Q | phys.abund.Fe | Fe/H abundance |
| Q | phys.abund.X | Hydrogen abundance |
| Q | phys.abund.Y | Helium abundance |
| Q | phys.abund.Z | Metallicity abundance |
| Q | phys.acceleration | Acceleration |
| Q | phys.albedo | Albedo or reflectance |
| Q | phys.angArea | Angular area |
| Q | phys.angMomentum | Angular momentum |
| Q | phys.angSize | Angular size, width, diameter, dimension, extension, major minor axis, extraction radius |
| Q | phys.angSize.smajAxis | Angular size, extent or extension of semi-major axis |
| Q | phys.angSize.sminAxis | Angular size, extent or extension of semi-minor axis |
| Q | phys.area | Area (in linear units) |
| S | phys.atmol | Atomic and molecular physics |
| Q | phys.atmol.branchingRatio | Branching ratio |
| Q | phys.atmol.collStrength | Collisional strength |
| Q | phys.atmol.collisional | Related to collisions |
| Q | phys.atmol.configuration | Configuration |
| Q | phys.atmol.crossSection | Atomic / molecular cross-section |
| Q | phys.atmol.element | Element |
| Q | phys.atmol.excitation | Atomic molecular excitation parameter |
| Q | phys.atmol.final | Quantity refers to atomic/molecular final/ground state, level, ecc. |
| Q | phys.atmol.initial | Quantity refers to atomic/molecular initial state, level, ecc. |
| Q | phys.atmol.ionStage | Ion |
| S | phys.atmol.ionization | Related to ionization |
| Q | phys.atmol.lande | Lande factor |
| S | phys.atmol.level | Atomic level |
| Q | phys.atmol.lifetime | Lifetime of a level |
| Q | phys.atmol.lineShift | Line shifting coefficient |
| Q | phys.atmol.number | Atomic number Z |
| Q | phys.atmol.oscStrength | Oscillator strength |
| Q | phys.atmol.parity | Parity |
| Q | phys.atmol.qn | Atomic/molecular quantum number |
| Q | phys.atmol.radiationType | Type of radiation characterizing atomic lines (electric dipole/quadrupole, magnetic dipole) |
| Q | phys.atmol.symmetry | Type of nuclear spin symmetry |
| Q | phys.atmol.sWeight | Statistical weight |
| Q | phys.atmol.sWeight.nuclear | Statistical weight for nuclear spin states |
| Q | phys.atmol.term | Atomic term |
| Q | phys.atmol.transProb | Atomic transition probability, Einstein A coefficient |
| S | phys.atmol.transition | Transition between states |
| Q | phys.atmol.wOscStrength | Weighted oscillator strength |
| Q | phys.atmol.weight | Atomic weight |
| Q | phys.columnDensity | Column density |
| S | phys.composition | Quantities related to composition of objects |
| Q | phys.composition.massLightRatio | Mass to light ratio |
| Q | phys.composition.yield | Mass yield |
| S | phys.cosmology | Related to cosmology |
| Q | phys.damping | Generic damping quantities |
| Q | phys.density | Density (of mass, electron, ...) |
| Q | phys.dielectric | Complex dielectric function |
| Q | phys.dispMeasure | Dispersion measure |
| V | phys.electField | Electric field |
| S | phys.electron | Electron |
| Q | phys.electron.degen | Electron degeneracy parameter |
| Q | phys.emissMeasure | Emission measure |
| Q | phys.emissivity | Emissivity |
| Q | phys.energy | Energy |
| Q | phys.energy.density | Energy-density |
| Q | phys.entropy | Entropy |
| Q | phys.eos | Equation of state |
| Q | phys.excitParam | Excitation parameter U |
| Q | phys.gauntFactor | Gaunt factor/correction |
| Q | phys.gravity | Gravity |
| Q | phys.ionizParam | Ionization parameter |
| Q | phys.ionizParam.coll | Collisional ionization |
| Q | phys.ionizParam.rad | Radiative ionization |
| E | phys.luminosity | Luminosity |
| Q | phys.luminosity.fun | Luminosity function |
| E | phys.magAbs | Absolute magnitude |

```
Q | phys.magAbs.bol              | Bolometric absolute magnitude
V | phys.magField                | Magnetic field
Q | phys.mass                    | Mass
Q | phys.mass.loss               | Mass loss
Q | phys.mol                     | Molecular data
Q | phys.mol.dipole              | Molecular dipole
Q | phys.mol.dipole.electric     | Molecular electric dipole moment
Q | phys.mol.dipole.magnetic     | Molecular magnetic dipole moment
Q | phys.mol.dissociation        | Molecular dissociation
Q | phys.mol.formationHeat       | Formation heat for molecules
Q | phys.mol.quadrupole          | Molecular quadrupole
Q | phys.mol.quadrupole.electric | Molecular electric quadrupole moment
S | phys.mol.rotation            | Molecular rotation
S | phys.mol.vibration           | Molecular vibration
S | phys.particle.neutrino       | Related to neutrino
E | phys.polarization            | Polarization degree (or percentage)
Q | phys.polarization.circular   | Circular polarization
Q | phys.polarization.linear     | Linear polarization
Q | phys.polarization.rotMeasure | Rotation measure polarization
Q | phys.polarization.stokes     | Stokes polarization
Q | phys.pressure                | Pressure
Q | phys.recombination.coeff     | Recombination coefficient
Q | phys.refractIndex            | Refraction index
Q | phys.size                    | Linear size, length (not angular)
Q | phys.size.axisRatio          | Axis ratio (a/b) or (b/a)
Q | phys.size.diameter           | Linear diameter
Q | phys.size.radius             | Linear radius
Q | phys.size.smajAxis           | Linear semi major axis
Q | phys.size.sminAxis           | Linear semi minor axis
Q | phys.temperature             | Temperature
Q | phys.temperature.effective   | Effective temperature
Q | phys.temperature.electron    | Electron temperature
Q | phys.transmission            | Transmission (of filter, instrument, ...)
V | phys.veloc                   | Space velocity
Q | phys.veloc.ang               | Angular velocity
Q | phys.veloc.dispersion        | Velocity dispersion
Q | phys.veloc.escape            | Escape velocity
Q | phys.veloc.expansion         | Expansion velocity
Q | phys.veloc.microTurb         | Microturbulence velocity
Q | phys.veloc.orbital           | Orbital velocity
Q | phys.veloc.pulsat            | Pulsational velocity
Q | phys.veloc.rotat             | Rotational velocity
Q | phys.veloc.transverse        | Transverse / tangential velocity
Q | phys.virial                  | Related to virial quantities (mass, radius, ..)
Q | pos                          | Position and coordinates
Q | pos.angDistance              | Angular distance, elongation
Q | pos.angResolution            | Angular resolution
Q | pos.az                       | Position in alt-azimutal frame
Q | pos.az.alt                   | Alt-azimutal altitude
Q | pos.az.azi                   | Alt-azimutal azimut
Q | pos.az.zd                    | Alt-azimutal zenith distance
S | pos.barycenter               | Barycenter
S | pos.bodyrc                   | Body related coordinates
Q | pos.bodyrc.alt               | Body related coordinate (altitude on the body)
Q | pos.bodyrc.lat               | Body related coordinate (latitude on the body)
Q | pos.bodyrc.long              | Body related coordinate (longitude on the body)
```

| | | |
|---|---|---|
| S | pos.cartesian | Cartesian (rectangular) coordinates |
| Q | pos.cartesian.x | Cartesian coordinate along the x-axis |
| Q | pos.cartesian.y | Cartesian coordinate along the y-axis |
| Q | pos.cartesian.z | Cartesian coordinate along the z-axis |
| S | pos.cmb | Cosmic Microwave Background reference frame |
| Q | pos.dirCos | Direction cosine |
| V | pos.distance | Linear distance |
| S | pos.earth | Coordinates related to Earth |
| Q | pos.earth.altitude | Altitude, height on Earth above sea level |
| Q | pos.earth.lat | Latitude on Earth |
| Q | pos.earth.lon | Longitude on Earth |
| S | pos.ecliptic | Ecliptic coordinates |
| Q | pos.ecliptic.lat | Ecliptic latitude |
| Q | pos.ecliptic.lon | Ecliptic longitude |
| S | pos.eop | Earth orientation parameters |
| Q | pos.eop.nutation | Earth nutation |
| Q | pos.ephem | Ephemeris |
| S | pos.eq | Equatorial coordinates |
| Q | pos.eq.dec | Declination in equatorial coordinates |
| Q | pos.eq.ha | Hour-angle |
| Q | pos.eq.ra | Right ascension in equatorial coordinates |
| Q | pos.eq.spd | South polar distance in equatorial coordinates |
| S | pos.errorEllipse | Positional error ellipse |
| Q | pos.frame | Reference frame used for positions (FK5, ICRS,..) |
| S | pos.galactic | Galactic coordinates |
| Q | pos.galactic.lat | Latitude in galactic coordinates |
| Q | pos.galactic.lon | Longitude in galactic coordinates |
| S | pos.galactocentric | Galactocentric coordinate system |
| S | pos.geocentric | Geocentric coordinate system |
| Q | pos.healpix | Hierarchical Equal Area IsoLatitude Pixelization |
| S | pos.heliocentric | Heliocentric position coordinate (solar system bodies) |
| Q | pos.HTM | Hierarchical Triangular Mesh |
| S | pos.lambert | Lambert projection |
| S | pos.lg | Local Group reference frame |
| S | pos.lsr | Local Standard of Rest reference frame |
| Q | pos.lunar | Lunar coordinates |
| Q | pos.lunar.occult | Occultation by lunar limb |
| Q | pos.parallax | Parallax |
| Q | pos.parallax.dyn | Dynamical parallax |
| Q | pos.parallax.phot | Photometric parallax |
| Q | pos.parallax.spect | Spectroscopic parallax |
| Q | pos.parallax.trig | Trigonometric parallax |
| Q | pos.phaseAng | Phase angle, e.g. elongation of earth from sun as seen from a third cel. object |
| V | pos.pm | Proper motion |
| Q | pos.posAng | Position angle of a given vector |
| V | pos.precess | Precession (in equatorial coordinates) |
| S | pos.supergalactic | Supergalactic coordinates |
| Q | pos.supergalactic.lat | Latitude in supergalactic coordinates |
| Q | pos.supergalactic.lon | Longitude in supergalactic coordinates |
| P | pos.wcs | WCS keywords |
| P | pos.wcs.cdmatrix | WCS CDMATRIX |
| P | pos.wcs.crpix | WCS CRPIX |
| P | pos.wcs.crval | WCS CRVAL |
| P | pos.wcs.ctype | WCS CTYPE |
| P | pos.wcs.naxes | WCS NAXES |
| P | pos.wcs.naxis | WCS NAXIS |
| P | pos.wcs.scale | WCS scale or scale of an image |
| Q | spect | Spectroscopy |
| Q | spect.binSize | Spectral bin size |
| S | spect.continuum | Continuum spectrum |
| Q | spect.dopplerParam | Doppler parameter b |
| E | spect.dopplerVeloc | Radial velocity, derived from the shift of some spectral feature |
| E | spect.dopplerVeloc.opt | Radial velocity derived from a wavelength shift using the optical convention |
| E | spect.dopplerVeloc.radio | Radial velocity derived from a frequency shift using the radio convention |
| E | spect.index | Spectral index |
| S | spect.line | Spectral line |
| E | spect.line.asymmetry | Line asymmetry |
| E | spect.line.broad | Spectral line broadening |
| Q | spect.line.broad.Stark | Stark line broadening coefficient |
| E | spect.line.broad.Zeeman | Zeeman broadening |
| E | spect.line.eqWidth | Line equivalent width |
| E | spect.line.intensity | Line intensity |
| E | spect.line.profile | Line profile |
| Q | spect.line.strength | Spectral line strength S |
| E | spect.line.width | Spectral line fwhm |
| Q | spect.resolution | Spectral (or velocity) resolution |
| S | src | Observed source viewed on the sky |
| S | src.calib | Calibration source |
| S | src.calib.guideStar | Guide star |
| Q | src.class | Source classification (star, galaxy, cluster...) |
| Q | src.class.color | Color classification |
| Q | src.class.distance | Distance class e.g. Abell |
| Q | src.class.luminosity | Luminosity class |
| Q | src.class.richness | Richness class e.g. Abell |
| Q | src.class.starGalaxy | Star/galaxy discriminator, stellarity index |
| Q | src.class.struct | Structure classification e.g. Bautz-Morgan |
| Q | src.density | Density of sources |
| Q | src.ellipticity | Source ellipticity |
| Q | src.impactParam | Impact parameter |
| Q | src.morph | Morphology structure |
| Q | src.morph.param | Morphological parameter |
| Q | src.morph.scLength | Scale length for a galactic component (disc or bulge) |
| Q | src.morph.type | Hubble morphological type (galaxies) |
| S | src.net | Qualifier indicating that a quantity (e.g. flux) is background subtracted rather than total |
| Q | src.orbital | Orbital parameters |
| Q | src.orbital.eccentricity | Orbit eccentricity |
| Q | src.orbital.inclination | Orbit inclination |
| Q | src.orbital.meanAnomaly | Orbit mean anomaly |
| Q | src.orbital.meanMotion | Mean motion |
| Q | src.orbital.node | Ascending node |
| Q | src.orbital.periastron | Periastron |
| Q | src.redshift | Redshift |
| Q | src.redshift.phot | Photometric redshift |
| Q | src.sample | Sample |
| Q | src.spType | Spectral type MK |
| Q | src.var | Variability of source |
| E | src.var.amplitude | Amplitude of variation |
| Q | src.var.index | Variability index |
| Q | src.var.pulse | Pulse |
| Q | stat | Statistical parameters |
| Q | stat.Fourier | Fourier coefficient |
| Q | stat.Fourier.amplitude | Amplitude Fourier coefficient |
| P | stat.correlation | Correlation between two parameters |
| P | stat.covariance | Covariance between two parameters |
| P | stat.error | Statistical error |
| P | stat.error.sys | Systematic error |
| Q | stat.filling | Filling factor (volume, time, ..) |
| Q | stat.fit | Fit |
| P | stat.fit.chi2 | Chi2 |
| P | stat.fit.dof | Degrees of freedom |
| P | stat.fit.goodness | Goodness or significance of fit |
| S | stat.fit.omc | Observed minus computed |
| Q | stat.fit.param | Parameter of fit |
| P | stat.fit.residual | Residual fit |
| P | stat.likelihood | Likelihood |
| S | stat.max | Maximum or upper limit |
| S | stat.mean | Mean, average value |
| S | stat.median | Median value |
| S | stat.min | Minimum or lowest limit |
| Q | stat.param | Generic statistical arameter |
| Q | stat.probability | Probability |
| P | stat.snr | Signal to noise ratio |
| P | stat.stdev | Standard deviation |
| S | stat.uncalib | Qualifier of a generic incalibrated quantity |
| Q | stat.value | Miscellaneous statistical value |
| P | stat.variance | Variance |
| P | stat.weight | Statistical weight |
| Q | time | Time, generic quantity in units of time or date |
| Q | time.age | Age |
| Q | time.creation | Creation time/date (of dataset, file, catalogue,...) |
| Q | time.crossing | Crossing time |
| Q | time.duration | Interval of time describing the duration of a generic event or phenomenon |
| Q | time.end | End time/date of a generic event |
| Q | time.epoch | Instant of time related to a generic event (epoch, date, Julian date, time stamp/tag,...) |
| Q | time.equinox | Equinox |
| Q | time.interval | Time interval, time-bin, time elapsed between two events, not the duration of an event |
| Q | time.lifetime | Lifetime |
| Q | time.period | Period, interval of time between the recurrence of phases in a periodic phenomenon |
| Q | time.phase | Phase, position within a period |
| Q | time.processing | A time/date associated with the processing of data |
| Q | time.publiYear | Publication year |
| Q | time.relax | Relaxation time |
| Q | time.release | The time/date data is available to the public |
| Q | time.resolution | Time resolution |
| Q | time.scale | Timescale |
| Q | time.start | Start time/date of generic event |

# Appendix B: Changes from previous versions

## Changes from v1.22

Text of par. 1.1 (2), last three lines;
List of em bands reordered according to wl/freq.

### Amendments/clarifications:

***Description*** changed in words:
phys.atmol.qn

### Additions:

em.line.Hdelta, em.line.Lyalpha, em.line.CO

### Deletions/replacements:

deleted: phys.mol.qn    replaced by: phys.atmol.qn

## Changes from v1.21

### Amendments/clarifications:

***Syntax flag*** changed in words:
phys.polarization

***Description*** changed in words:
em.IR.FIR, em.IR.MIR, em.IR.NIR, em.line.OIII

## Changes from v1.2

### Additions:

spect.continuum

## Changes from v1.11 (Rec20051231)

### Amendments/clarifications:

***Spelling***: phys.atmol.sWeight

***Syntax flag*** changed in words:
phys.atmol, spect.line

***Description*** changed in words:
meta.dataset, obs.atmos, phot.color.reddFree, phys.size, phys.size.diameter, phys.size.radius, stat.param, stat.value, time, time.epoch, time.interval, time.period, time.phase

### Additions:

em.bin, em.binSize, em.IR.FIR, em.IR.MIR, em.IR.NIR, em.UV.FUV, meta.abstract, meta.code.status, meta.email, meta.id.PI, meta.id.CoI, meta.ref.ivorn, meta.ref.uri, obs.calib.flat, obs.exposure, obs.proposal, obs.proposal.cycle, obs.sequence, phys.atmol.symmetry, phys.atmol.sWeight.nuclear, phys.cosmology, phys.damping, phys.entropy, phys.particle.neutrino, phys.virial, spect.line.strength, src.calib, src.calib.guideStar, src.net, stat.filling, stat.probability, stat.uncalib, time.creation, time.duration, time.end, time.processing, time.publiYear, time.release, time.start

**Deletions/replacements:**

| deleted | replacement | description |
|---|---|---|
| phys.atmol.damping | phys.damping | Atomic damping quantities (van der Waals) |
| phys.atmol.qn.I | phys.atmol.qn | Nuclear spin quantum number |
| time.event | time.duration | Duration of an event or phenomenon |
| time.event.end | time.end | End time of event or phenomenon |
| time.event.start | time.start | Start time of event or phenomenon |
| time.expo | time.duration;obs.exposure | Exposure on-time, duration |
| time.expo.end | time.end;obs.exposure | End time of exposure |
| time.expo.start | time.start;obs.exposure | Start time of exposure |
| time.obs | time.duration;obs | Observation on-time, duration |
| time.obs.end | time.end;obs | End time of observation |
| time.obs.start | time.start;obs | Start time of observation |

## Changes from v1.10

1. A few minor changes to the text have been done
2. All UCD words are now compliant with the UCD recommendation. The corresponding changes are described below
3. The following words have been deprecated:

| Deprecated UCD | New corresponding UCD |
|---|---|
| phot.fluxDens | phot.flux.density |
| phot.fluxDens.sb | phot.flux.density.sb |
| **phys.at**\* | **phys.atmol**\* |
| phys.atmol.coll | phys.atmol.collisional |
| phys.atmol.ion | phys.atmol.ionStage |
| phys.atmol.trans | phys.atmol.transition |
| phys.energyDensity | phys.energy.density |
| phys.massToLight | phys.composition.massLightRatio |
| phys.massYield | phys.composition.yield |
| spect.doppler | spect.dopplerParam |

4. The following word has been created: phys.composition

5. The section *Changes from previous versions* has been reformatted

# Changes from v1.02

   1. Descriptions have been changed for the following words: em.line, instr.pixel, phys.gravity, pos.earth.altitude
   2. The syntax flags changed for words: instr.filter, phys.angSize
   3. The following words have been deprecated:

| Deprecated UCD | New corresponding UCD |
|---|---|
| instr.filter.transm | phys.transm;instr.filter |
| phys.mass.light | phys.massToLight |
| pos.resolution | pos.angResolution |
| pos.satellite | pos.bodyrc |

   4. The following words have been created: phys.polarization.circular, phys.polarization.linear, phys.size.axisRatio, pos.bodyrc.alt, pos.bodyrc.lat, pos.bodyrc.long, time.event, time.event.end, time.event.start

# Changes from v1.01

   1. The following words have been restored to their previous spelling (v1.00): phot.fluDensity, phys.energDensity, phys.mYield, phot.fluxDensity, phys.energyDensity, phys.massYield

      A note has been added to indicate that these words do not strictly comply with the UCD1+ Rec.

# Changes from v1.00

   1. Descriptions have been changed for the following words: em.IR.H, em.IR.J, em.IR.K, em.X-ray.hard, em.X-ray.medium, em.X-ray.soft, em.gamma.hard, em.gamma.soft, em.opt.B, em.opt.I, em.opt.R, em.opt.U, em.opt.V, instr.bandpass, phot.count, phys.density, phys.mol.dipole.electric, phys.mol.dipole.magnetic, phys.mol.quadrupole.electric, pos.angDistance, pos.precess, src, src.class.distance, src.class.richness, src.class.starGalaxy, src.class.struct, time.expo, time.expo.end, time.expo.start, time.interval
   2. The following words have been deprecated:

| Deprecated UCD | New corresponding UCD |
|---|---|
| instr.angRes | pos.resolution |
| instr.obsty.site | pos.earth.altitude;instr.obsty |
| instr.obsty.site.seeing | instr.obsty.seeing |
| instr.spect | instr |
| instr.spect.dispersion | instr.dispersion |
| instr.spect.order | instr.order |
| instr.spect.resolution | spect.resolution |
| instr.tel.focus | instr.tel.focalLength |
| meta.fits.software | meta.software |
| obs.air | obs.atmos |
| obs.air.extinction | obs.atmos.extinction |
| obs.air.mass | obs.airMass |
| phot.fluxDens | phot.fluDens |
| phot.fluxDens.sb | phot.fluDens.sb |
| phot.sb | phot.mag.sb |
| phys.at.branchingRatio | phys.atmol.branchingRatio |
| phys.at.crossSection | phys.atmol.crossSection |
| phys.at.lineShift | phys.atmol.lineShift |
| phys.at.moment | |
| phys.at.moment.electric | phys.at.radiationType |
| phys.at.moment.magnetic | phys.at.radiationType |
| phys.at.qn.S | phys.at.qn |
| phys.at.qn.L | phys.at.qn |
| phys.at.qn.J | phys.at.qn |
| phys.at.qn.F | phys.at.qn |
| phys.atmol.state.final | phys.atmol.final |
| phys.atmol.state.initial | phys.atmol.initial |
| phys.massYield | phys.mYield |
| phys.mol.quadrupole.magnetic | phys.at.radiationType |
| phys.refraction | phys.refractIndex |
| pos.az.ha | pos.eq.ha |
| pos.earth.nutation | pos.eop.nutation |
| spect.veloc | spect.dopplerVeloc |
| src.fwhm | phys.angSize;src |
| src.orbital.veloc | phys.veloc.orbital |
| src.veloc | phys.veloc |
| src.veloc.ang | phys.veloc.ang |
| src.veloc.cmb | phys.veloc;pos.cmb |
| src.veloc.dispersion | phys.veloc.dispersion |
| src.veloc.escape | phys.veloc.escape |
| src.veloc.expansion | phys.veloc.expansion |
| src.veloc.lg | phys.veloc;pos.lg |
| src.veloc.lsr | phys.veloc;pos.lsr |
| src.veloc.microTurb | phys.veloc.microTurb |
| src.veloc.pulsat | phys.veloc.pulsat |
| src.veloc.rotat | phys.veloc.rotat |

3. The syntax flags changed for words: instr.fov, instr.obsty, meta.file, phys.angSize, pos.cartesian, stat.fit.omc
4. The following words have been created: instr.dispersion, instr.order, instr.tel.focalLength, meta.curation, meta.software, meta.version, obs.atmos, obs.atmos.extinction, obs.airMass, obs.atmos.refractAngle, obs.calib, phys.at.radiationType, phys.atmol.branchingRatio, phys.atmol.crossSection, phys.atmol.lifetime, phys.atmol.lineShift, phys.energDensity, phys.refractIndex, phys.transmission, pos.eq.ha, pos.az.azi, pos.bodyrc, pos.cmb, pos.earth.altitude, pos.eop, pos.eop.nutation, pos.lg, pos.lsr, pos.phaseAng, pos.resolution, spect.resolution, spect.dopplerVeloc, spect.dopplerVeloc.radio, spect.dopplerVeloc.opt,

src.orbital.meanMotion, phys.veloc, phys.veloc.ang, phys.veloc.dispersion, phys.veloc.escape, phys.veloc.expansion, phys.veloc.microTurb, phys.veloc.orbital, phys.veloc.pulsat, phys.veloc.rotat, phys.veloc.transverse, time.obs, time.obs.end, time.obs.start

## Changes from v0.2

1. Section 1.2 has been simplified
2. 3 new syntax codes (E, C, V) have been introduced, and described in appendix A
3. The following words have been renamed :

| Deprecated UCD | New corresponding UCD |
|---|---|
| em.line.21cm | em.line.HI |
| instr.ang-res | instr.angRes |
| instr.sky-level | instr.skyLevel |
| instr.sky-temp | instr.skyTemp |
| instr.antenna-temp | phot.antennaTemp |
| phys.absorption.gf | phys.gauntFactor |
| phys.at.einstein | phys.at.transProb |
| phys.at.level | phys.atmol.level |
| phys.dispMeas | phys.dispMeasure |
| phys.distance | pos.distance |
| phys.polarization.rotMeas | phys.polarization.rotMeasure |
| phys.size.area | phys.area |
| pos.ang.separation | pos.angDistance |
| pos.ec | pos.ecliptic |
| pos.ec.lat | pos.ecliptic.lat |
| pos.ec.lon | pos.ecliptic.lon |
| pos.ee | pos.errorEllipse |
| pos.gal | pos.galactic |
| pos.gal.lat | pos.galactic.lat |
| pos.gal.lon | pos.galactic.lon |
| pos.sg | pos.supergalactic |
| pos.sg.lat | pos.supergalactic.lat |
| pos.sg.lon | pos.supergalactic.lon |
| src.class.star-galaxy | src.class.starGalaxy |

4. The following words have been created: instr.beam, meta.code.error, meta.id.part, phot.flux.sb, phys.angArea, phys.angSize, phys.angSize.smajAxis, phys.angSize.sminAxis, phys.area, phys.at.damping, phys.at.weight, phys.atmol.excitation, phys.mol.dissociation, phys.recombination.coeff, phys.size.smajAxis, phys.size.sminAxis, pos.cartesian, pos.cartesian.x, pos.cartesian.y, pos.cartesian.z, pos.distance, pos.eq.spd, pos.galactocentric, pos.geocentric, pos.healpix, pos.heliocentric, pos.HTM, pos.lambert, pos.satellite, spect.line.broad.Stark, spect.veloc, src.redshift.phot, stat.correlation, time.lifetime
5. Some words have been removed. The following table summarizes, when relevant, the suggested replacement to be used.

| Deprecated UCD | New corresponding UCD |
|---|---|
| instr.area | phys.area;instr |
| instr.beam-width | phys.angSize;instr.beam |
| meta.table.axis | phys.size;meta.table |
| phot.color.Cous | phot.color |
| phot.color.Gen | phot.color |
| phot.color.Gunn | phot.color |
| phot.color.JHN | phot.color |
| phot.color.STR | phot.color |
| phot.color.STR.c1 | phot.color |
| phot.color.STR.b-y | phot.color |
| phot.color.STR.m1 | phot.color |
| phys.at.lineBroad | spect.line.broad |
| phys.distance.compon | pos.distance;pos.cartesian.x (or y, z) |
| phys.distance.gc | pos.distance;pos.galactocentric |
| phys.electron.energy | phys.energy;phys.electron |
| phys.extension | phys.angSize or phys.size |
| phys.mass.fraction | phys.mass;arith.ratio |
| phys.polarization.posAng | pos.posAng;phys.polarization |
| pos.ang | |
| pos.det | pos.cartesian;instr.det |
| pos.eq.dec.arcsec | |
| pos.eq.ra.minutes | |
| pos.eq.ra.seconds | |
| pos.gal.compon | pos.cartesian;pos.galactic |
| pos.pm.dec | pos.pm;pos.eq.dec |
| pos.pm.ra | pos.pm;pos.eq.ra |
| pos.precess.dec | pos.precess;pos.eq.dec |
| pos.precess.ra | pos.precess;pos.eq.ra |
| pos.proj | |
| pos.sg.compon | pos.cartesian;pos.supergalactic |
| src.orbital.energy | phys.energy;src.orbital |
| src.orbital.separation | pos.angDistance;src.orbital |
| src.orbital.size | phys.size;src.orbital |
| src.separation | pos.angDistance;src |
| src.veloc.compon | src.veloc;pos.cartesian |
| src.veloc.gc | src.veloc;pos.galactocentric |
| src.veloc.geoc | src.veloc;pos.geocentric |
| src.veloc.hc | src.veloc;pos.heliocentric |

## Changes from v0.1

1. Descriptions of the words were improved.
2. Designation of commonly used lines have been moved to **em.line.***. As a consequence, terms like **em.IR.K.Brgamma** or **spect.index.Hbeta** have been removed.

3. **phys.at** and **phys.mol** have been completely reorganized to improve the overall description of this domain. A new branch **phys.atmol** has been introduced to group concepts shared between **phys.at** and **phys.mol**.
4. The **phot.color** section was significantly simplified.
5. Missing nodes of the tree were added (e.g. **em.gamma**, **em.mm**, **pos.sg**).
6. Creation of new words: **em.wavenumber**, **meta.ucd**, **stat.error.sys**
7. Typos were corrected in **em.opt.*** units and a few other descriptions.

# References

[1] R. Hanisch, *Resource Metadata for the Virtual Observatory* , http://www.ivoa.net/Documents/latest/RM.html

[2] R. Hanisch, M. Dolensky, M. Leoni, *Document Standards Management: Guidelines and Procedure* , http://www.ivoa.net/Documents/latest/DocStdProc.html